\documentclass[
   ,final            
  ]
  {aipproc}

\layoutstyle{8x11double}

  \def\lsim{\buildrel < \over {_{\sim}}}

\begin{document}

\title{Overview of Neutrino Mixing Models and\\
       Their Mixing Angle Predictions\footnote{Written version of a talk
       presented at the 11th International Workshop on Neutrino Factories,
       Superbeams and Beta Beams, Illinois Institute of Technology, Chicago,
       July 20-25, 2009.}}


\author{Carl H. Albright}{
  address={Department of Physics, Northern Illinois University, DeKalb, IL 60115, USA}
  ,altaddress={Theoretical Physics Department, Fermi National Accelerator Laboratory,
  Batavia, IL 60510, USA}  
}



\begin{abstract}
  An overview of neutrino-mixing models is presented 
  with emphasis on the types of horizontal flavor and vertical 
  family symmetries that have been invoked. Distributions for
  the mixing angles of many models are displayed.
  Ways to differentiate among the models and to narrow the 
  list of viable models are discussed.

\end{abstract}

\maketitle


\section{Introduction}
  Several hundred models of neutrino masses and mixings
  can be found in the literature which purport to explain the 
  known oscillation data and predict the currently unknown
  quantities.  We present an overview of the types of models
  proposed and discuss ways in which the list of viable 
  models can be reduced when more precise data is obtained. 
  This presentation is an update of one published in 2006 in
  collaboration with Mu-Chun Chen \cite{a-c} and, due to space
  restrictions, is an abreviated version of one appearing in 
  \cite{cha} with complete model references there.
\vspace*{-0.1in}
\section{Present Oscillation Data and Unknowns}
  The present data within $3\sigma$ accuracy as determined by 
  Fogli et al. \cite{fogli}, for example, is given by 
  \begin{eqnarray}
   \label{eq:data} 
    \Delta m^2_{32} &=& 2.39\begin{array}{c} +0.42 \\ -0.33 \\ \end{array}
    \times 10^{-3}\ {\rm eV^2,} \nonumber \\
     \Delta m^2_{21} &=& 7.67\begin{array}{c} +0.52\\ -0.53\\ \end{array}
    \times 10^{-5}\ {\rm eV^2}, \nonumber \\
    \sin^2 \theta_{23} &=& 0.466\begin{array}{c} +0.178\\ -0.135\\ \end{array},
      \nonumber \\
    \sin^2 \theta_{12} &=& 0.312\begin{array}{c} +0.063 \\ -0.049\\ 
      \end{array}, \nonumber \\
    \sin^2 \theta_{13} &\leq& 0.046, \quad (0.016 \pm 0.010), 
   \end{eqnarray}

\noindent  where the last figure in parenthesis indicates a departure 
of the reactor neutrino angle from zero with one $\sigma$
accuracy determination.
The data suggests the approximate tri-bimaximal 
mixing texture of Harrison, Perkins, and Scott \cite{tbm},
\begin{equation}
  U_{PMNS} = \left( \begin{array}{ccc} 2/\sqrt{6} & 1/\sqrt{3} & 0\\
             -1/\sqrt{6} & 1/\sqrt{3} & -1/\sqrt{2}\\
	   -1/\sqrt{6} & 1/\sqrt{3} & 1/\sqrt{2} \end{array}\right),
\label{eq:PMNS}
\end{equation}

\noindent with $\sin^2 \theta_{23} = 0.5,\ \sin^2 \theta_{12} = 0.33$,
and $\sin^2 \theta_{13} = 0$.

The reason for the plethera of models still in agreement with experiment
of course can be traced to the inaccuracy of the present data and the 
imprecision of the model predictions in many cases.  In 
addition, there are a number of unknowns that must still be determined:  
the hierarchy and absolute mass scales of the light neutrinos; the Dirac 
or Majorana nature of the neutrinos; the CP-violating phases of the mixing 
matrix; how close to zero the reactor neutrino angle, $\theta_{13}$, lies;
how near maximal the atmospheric neutrino mixing angle is; whether the 
approximate tri-bimaximal mixing is a softly-broken or an accidental symmmetry;
whether neutrino-less double beta decay will be observable, and how large 
charged lepton flavor violation will turn out to be.
In this presentation we survey the models to determine what they predict
for the mixing angles, and neutrino mass hierarchy. 
\vspace*{-0.1in}
\section{Theoretical Framework}
The observation of neutrino oscillations implies that neutrinos have mass,
with the mass squared differences given in Eq.(\ref{eq:data}).  Information
concerning the absolute neutrino mass scale has been determined by the 
combined WMAP, SDSS, and Lyman alpha data which place an upper limit on 
the sum of the masses \cite{masssum},
\begin{equation}
  \sum_i m_i \leq 0.17 - 1.2\ {\rm eV},
\label{eq:masssum}
\end{equation}

\noindent depending upon the conservative nature of the bound extracted.
An extension of the SM is then required, and possible approaches 
include one or more of the following:
\begin{itemize}
\item the introduction of dim-5 effective non-renormalizable operators;
\item the addition of right-handed neutrinos with their Yukawa couplings
  to the left-handed neutrinos;
\item the addition of direct mass terms with right-handed Majorana couplings;
\item the addition of a Higgs triplet with left-handed Majorana couplings;
\item the addition of a fermion triplet with Higgs doublet couplings.
\end{itemize}

If we exclude the last possibility, the general $6 \times 6$ neutrino mass 
matrix in the $B(\nu_{\alpha L},\ N^c_{\alpha L})$ flavor basis of the six 
left-handed fields then has the following structure in terms of $3 \times 3$ 
submatrices:
\begin{equation}
  \cal{M} = \left(\begin{array}{cc} M_L & M^T_N \\ M_N & M_R \end{array}
             \right),
\label{eq:6x6}
\end{equation}

\noindent where $M_N$ is the Dirac neutrino mass matrix, $M_L$ the 
left-handed and $M_R$ the right-handed Majorana neutrino mass matrices.
With $M_L = 0$ and $M_N << M_R$ the type~I seesaw formula,  
\begin{equation}
  m_\nu = - M^T_N M^{-1}_R M_N,
\label{eq:typeI}
\end{equation}

\noindent is obtained for the light left-handed Majorana neutrinos, while if 
$M_L \neq 0$ and $M_N << M_R$, one obtains the mixed type I + II seesaw formula,
\begin{equation}
  m_\nu = M_L - M^T_N M^{-1}_R M_N.
\label{eq:typeII}
\end{equation}

There are two main approaches which we now describe that one can 
pursue to learn more about the theory behind the lepton mass generation.

\subsection{Top - Down Approach}
In the top-down approach one postulates the form of the mass matrix from
first principles.  The models will differ then due to the horizontal flavor 
symmetry chosen, the vertical family symmetry (if any) selected, and the 
fermion and Higgs representation assignments made.

The effective light left-handed Majorana mass matrix $m_\nu$ is constructed 
directly or with the seesaw formula once the Dirac neutrino matrix $M_N$ 
and the Majorana neutrino matrices $M_R$ (and $M_L$) are specified.
Since $m_\nu$ is complex symmetric, it can be diagonalized by a unitary 
transformation, $U_{\nu_L}$, to give
\begin{equation}
  m^{diag}_\nu = U^T_{\nu_L} m_\nu U_{\nu_L} = {\rm diag}(m_1,\ m_2,\ m_3),
\end{equation}

\noindent with real, positive masses down the diagonal.  On the other hand,
the Dirac charged lepton mass matrix is diagonalized by a bi-unitary 
transformation according to 
\begin{equation}
  m^{diag}_\ell = U^\dagger_{\ell R} m_\ell U_{\ell L} = 
  {\rm diag}(m_e,\ m_\mu, m_\tau).
\end{equation}

\noindent The neutrino mixing matrix \cite{PMNS}, $V_{PMNS}$, is then given 
by \\
\begin{equation}
  V_{PMNS} \equiv U^\dagger _{\ell L} U_{\nu_L} = U_{PMNS}\Phi,\\
\label{eq:VPMNS}
\end{equation}

\noindent   in the lepton flavor basis with $\Phi = {\rm diag}(1,e^{i\alpha},
\ e^{i\beta})$.  Note that the Majorana phase matrix 
$\Phi$ is required in order to compensate for any phase rotation on 
$U_{\nu_L}$ needed to bring it into the Particle Data Book phase 
convention \cite{pdb}.
  
\subsection{Bottom - Up Approach}
On the other hand, with a bottom-up approach in the diagonal lepton flavor
basis and with the general PMNS mixing matrix, one can determine 
the general texture of the light neutrino mass matrix to be 
\begin{eqnarray}
\label{eq:bot-up}
 M_\nu &=& U^*_{PMNS}\Phi^* M^{\rm diag}_\nu \Phi^*U^\dagger_{PMNS}\nonumber\\
	&=& U^*_{PMNS}{\rm diag}(m_1,
\ m_2 e^{-2i\alpha},\ m_3 e^{-2i\beta})U^\dagger_{PMNS}\nonumber\\[0.1in]
&\equiv& \left(\begin{array}{ccc}
A & B & B'\\ \cdot & F' & E\\ \cdot & \cdot & F\\ \end{array}\right),
\end{eqnarray}

\noindent  where the matrix elements are expressed in terms of the unknown
neutrino masses, mixing angles and phases.  By restricting the mixing matrix,
one can learn that some of the matrix elements may not be independent.
  
\section{Models and Mixing Angle Predictions}
After suggestions of atmospheric neutrino oscillations were found by the  
IMB and Kamiokande-II Collaborations \cite{atm} in the early 1990's, it 
became fashionable to assign texture zeros in different positions to 
$m_\nu$ with a top-down approach in hopes of identifying some flavor 
symmetry, but the procedure is basis dependent \cite{textzero}. 

Another popular method invoked a $L_e - L_\mu - L_\tau$ lepton flavor
symmetry \cite{e-mu-tau}.  The mass matrix then assumes the following form
\begin{equation}
  m_\nu = \left(\begin{array}{ccc} 0 & * & * \\ \cdot & 0 & 0 \\ 
    \cdot & 0 & 0 \\ 
  \end{array}
  \right), \\
\end{equation}

\noindent which only leads to an inverted hierarchy.

By making use of a bottom-up approach instead, one is able to observe that a 
$\mu~-~\tau$ interchange symmetry with $B' = B,\ F' = F$ in 
Eq. (\ref{eq:bot-up}) leads to $\sin^2 \theta_{23} = 0.5,\ \sin^2 \theta_{13} 
= 0$ with $\sin^2 \theta_{12}$ arbitrary.  

On the other hand, with the assumption of exact tri-bimaximal mixing for
which $\sin^2 \theta_{23} = 0.5,\ \sin^2 \theta_{13} = 0$, and $\sin^2 
  \theta_{12} = 0.333$, one finds in Eq. (\ref{eq:bot-up}) that $B' = B,
  \ F' = F = \frac{1}{2}(A + B + D)$ and $E = \frac{1}{2}(A + B - D)$, so 
that just three unknowns are present.

With the realization in the past five years that neutrino mixing is well 
approximated by the tri-bimaximal mixing matrix, the name of the game 
has become one of finding what discrete horizontal flavor symmetry groups 
would lead naturally to this mixing pattern.  Such flavor symmetries can 
then be used as starting points with soft breaking as the next approximation.
  
\subsection{Discrete Horizontal Flavor Symmetry Groups}
Of special interest are those groups containing doublet and triplet
irreducible representations.   We list several of the well-studied groups
and pertinent features of each.

The permutation group of three objects, $S_3$, contains 6 elements with 
$1, 1',$ and 2 dimensional irreducible representations (IR's).  The same 
eigenstates occur as those for tri-bimaximal mixing, but there is a 2-fold 
neutrino mass degeneracy.

The group $A_4$ of even permutations of four objects has 12 elements with 
IR's labeled $1, 1', 1''$, and 3.  A $U(1)_R$ symmetry \cite{f-n} may also 
be included to fix the mass scale which is otherwise undetermined.
Early attempts to extend this flavor group to the quark sector failed, as 
the CKM mixing matrix for the quarks remained diagonal.

The group $T'$ is the covering group of $A_4$, but interestingly $A_4$ 
is not one of its subgroups.  It contains 24 elements with $1, 1', 1'', 3,
2, 2', 2''$ IR's, where the first four are identical to those in $A_4$.   
While tri-bimaximal mixing is obtained for the leptons, due to the presence 
of the three doublet IR's, a satisfactory CKM mixing matrix can also be 
obtained for the quarks.

The permutation group of 4 objects, $S_4$, has 24 elements with 
$1, 1',2, 3, 3'$ IR's.  Although this and higher dimensional discrete 
flavor groups can also yield tri-bimaximal mixing, it appears that models
based on $A_4$ are the most economical ones for the lepton sector.

\subsection{Examples Involving GUT Models}
Studies of neutrino mixing models in the framework of grand unified 
theories with a vertical family symmetry were first pursued in the 1990's 
and more intensely following the discovery of atmospheric neutrino 
oscillations by the Super-Kamiokande Collaboration in 1998.  Examples exist
of models based on $SU(5)$, $SO(10)$, and $E_6$, where the 
$SO(10)$ models are generally of two types.

The so-called ``minimal'' $SO(10)$ models \cite{minimal} involve Higgs fields 
appearing in the ${\bf 10}$ and ${\bf 126}$ IR's, but newer models of this 
type have been extended to include the ${\bf 120}$, ${\bf 45}$, and/or 
${\bf 54}$ IR's.
They generally result in symmetric and/or antisymmetric contributions
to the quark and lepton mass matrices.

On the other hand, $SO(10)$ models \cite{lopsided} with Higgs fields in the 
${\bf 10, 16, \overline{16}}$ and ${\bf 45}$ IR's result in ``lopsided''
down quark and charged lepton mass matrices due to the $SU(5)$
structure of the electroweak VEV's appearing in the ${\bf 16}$ and 
$\overline{\bf 16}$ representations.

For either type of GUT model, type I seesaws only lead to a stable
normal hierarchy for the light neutrino masses \cite{alb}, while type I + II 
seesaws can also result in an inverted hierarchy.  Most of the 
$SO(10)$ models have a continuous and/or discrete flavor 
symmetry group producted with them, but no efforts were initially made to 
introduce a discrete flavor symmetry group from one of the types discussed
earlier to achieve tri-bimaximal mixing.  A few examples can now be found 
in the literature which combine an $SU(5),\ SO(10)$ or $E_6$ GUT symmetry 
with a $T'$ or $A_4$ flavor symmetry with some success \cite{famflav}.

\section{Survey of Mixing Angle and Hierarchy Predictions}

The author has updated a previous survey \cite{a-c} made in collaboration
with Mu-Chun Chen in 2006 of models in the literature which satisfied
the then current experimental bounds on the mixing angles and gave
reasonably restrictive predictions for the reactor neutrino angle.
The cutoff date for the present update is January 2009.

Many models in the literature lack firm predictions for any of the mixing 
angles.  For our analysis no requirement is 
made that the solar and atmospheric mixing angles or the mass differences
be predicted, but if so, they must also satisfy the bounds given in 
Eq. (\ref{eq:data}).  A complete listing of the 86 models which meet our 
criteria are referenced along with their predictions in \cite{cha}.

Here we simply present the model predictions in the form of histograms 
plotted against $\sin^2 \theta_{13}$, where all models are assigned the 
same area, even if they extend across several basic intervals.  
The results are shown in Figs. 1 and 2 for the lepton flavor 
models and grand unified models, respectively.  Two thirds of
both types of models predict $0.001 \lsim \sin^2 \theta_{13} \lsim 0.05$,
while the lepton flavor models have a much longer tail extending
to very small reactor neutrino angles.  The planned  
experiments involving Double Chooz and Daya Bay reactors \cite{reac}
will reach down to $\sin^2 2\theta_{13} \lsim 0.01$, so roughly 
two-thirds of the models will be eliminated if no $\bar{\nu}_e$ 
depletion is observed.  Both the T2K Collaboration at JPARC and 
the NO$\nu$A Collaboration at Fermilab are also expected to probe
a similar reach with their $\nu_\mu$ neutrino beams \cite{T2KNOVA}.

Even if $\bar{\nu}_e$ depletion is observed with some accuracy,
it is apparent from the two histograms that the order of 10 - 20
models may survive which must still be differentiated.  One 
suggestion is to make scatterplots of $\sin^2 \theta_{13}\ vs.\ 
\sin^2 \theta_{12}$ and $\sin^2 \theta_{12}\ vs.\ 
\sin^2 \theta_{23}$.  We have attempted to do this in Figs.
3, 4, and 5 for both the lepton flavor models and grand unified
models, where only the central value predictions are plotted.
Most of the models considered favor central values of 
$\sin^2 \theta_{12}$ lying below 0.333, the value for exact
tri-bimaximal mixing.  This is in agreement with the present
value extracted in Eq. (\ref{eq:data}), but central values for 
$\sin^2 \theta_{23} \geq 0.5$ are preferred, while the best extracted 
value is 0.466 from Eq. (\ref{eq:data}).

\begin{figure}[t]
  \includegraphics[width=7.9cm,height=6cm]{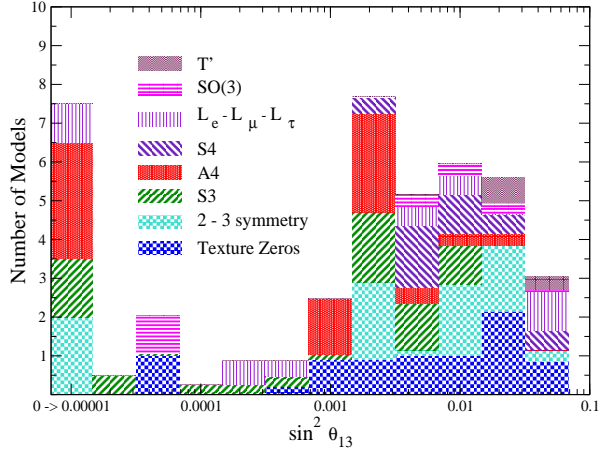}
  \caption{Lepton flavor model predictions for $\sin^2 \theta_{13}$.}
\label{fig:LFmodels}
\end{figure}
\vspace*{1cm}
\begin{figure}[h]
  \includegraphics[width=7.9cm,height=6cm]{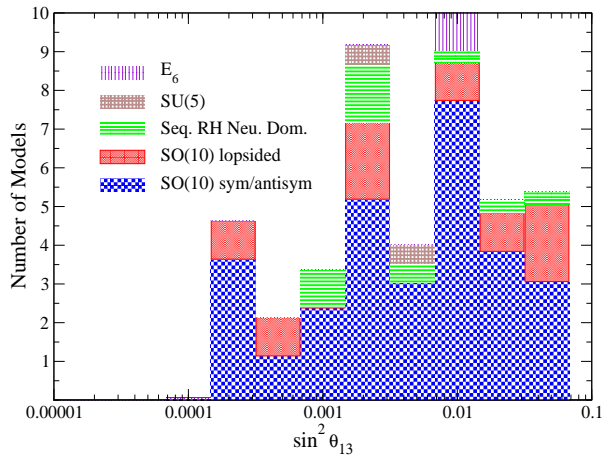}
  \caption{GUT model predictions for $\sin^2 \theta_{13}$.}
\label{fig:GUTmodels}
\end{figure}

\begin{figure}[t]
  \includegraphics[width=7.5cm,height=5.4cm]{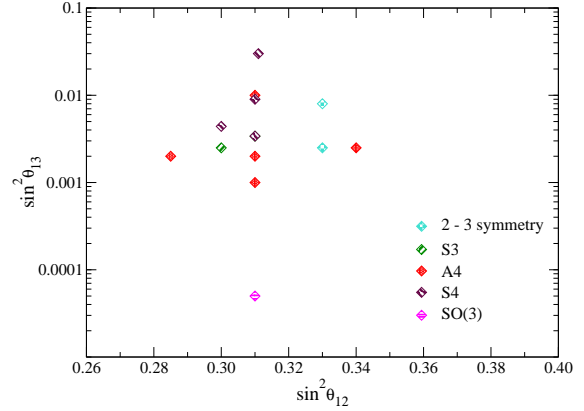}
  \caption{Central value distributions of $\sin^2 \theta_{23}\ vs.$  
  $\sin^2 \theta_{12}$ for the discrete flavor symmetry models.}
\label{fig:LFM1213}
\end{figure}
\vspace*{0.37cm}
\begin{figure}[!h]
  \includegraphics[width=7.5cm,height=5.4cm]{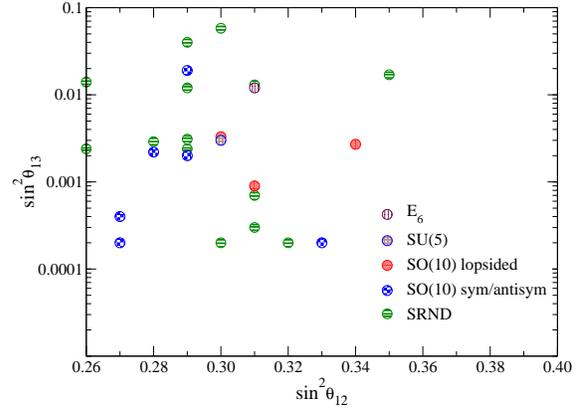}
  \caption{Central value distributions of $\sin^2 \theta_{23}\ vs.$ 
  $\sin^2 \theta_{12}$ for the grand unified models.}
\label{fig:GUT1213}
\end{figure}
\vspace*{0.65cm}
\begin{figure}[!h]
  \includegraphics[width=7.2cm,height=5.4cm]{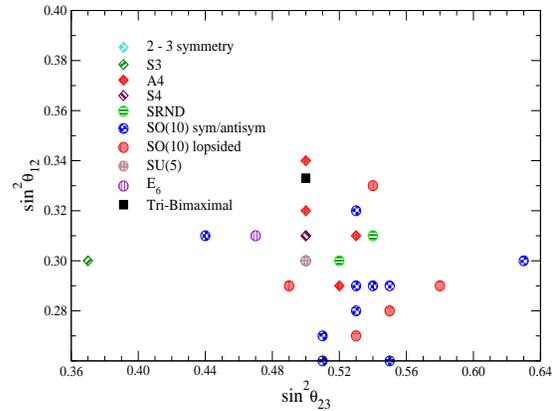}
  \caption{Central value distributions of $\sin^2 \theta_{12}\ vs.$ 
  $\sin^2 \theta_{23}$ for both types of models.}
\label{fig:LFMGUT2312}
\end{figure}

Neutrino-less double beta decay can serve as a valuable probe 
of the neutrino mass hierarchy observed in Nature.  In fact, the
effective mass plot for perturbed tri-bimaximal mixing in Fig. 6 shows
a clear separation of the normal and inverted ordering 
distributions, so accurate neutrino-less double beta decay
experiments should be decisive in determining the hierarchy.

\section{Summary} 
We have made a survey of neutrino mixing models based on  
some horizontal lepton flavor symmetry and those based on GUT models 
having a vertical family symmetry and a flavor symmetry.  We have tried to 
differentiate the models on the basis of their neutrino mass hierarchy and
mixing angles.  Most of the models allow either mass hierarchy with the 
exceptions being just normal for the type I seesaw models and only inverted
for the conserved $L_e - L_\mu - L_\tau$ models.

For both types of models our study indicates that the upcoming 
Double Chooz and Daya Bay reactor experiments will be able 
to eliminate roughly two-thirds of the models surveyed, if their 
planned sensitivity reaches $\sin^2 2\theta_{13} \simeq  0.01$ 
and no depletion of the $\bar{\nu}_e$ flux is observed.  However,
no smoking gun apparently exists to rule out many types of models
based on accurate data for $\sin^2 \theta_{13}$ alone, should 
evidence for a depletion be found.  Of the order of 10 - 20 models
have similar values for this mixing angle in the 0.001 - 0.05
interval.   Only the lepton flavor models appear to lead to 
extremely small values of $\sin^2 \theta_{13} \lsim 10^{-4}$.

Most models prefer $\sin^2 \theta_{12} \lsim 0.31$ rather than
0.333 for tri-bimaximal mixing in agreement with the present 
best value of 0.312.  On the other hand, most models prefer
$\sin^2 \theta_{23} \geq 0.50$ compared with a best fit 
value of 0.466.  One notable exception is the model of Stech and 
Tavartkiladze \cite{st} based on the $E_6$ family group with 
$SU(3) \times Z_2$ flavor symmetry.

\begin{figure}
  \includegraphics[width=7.5cm]{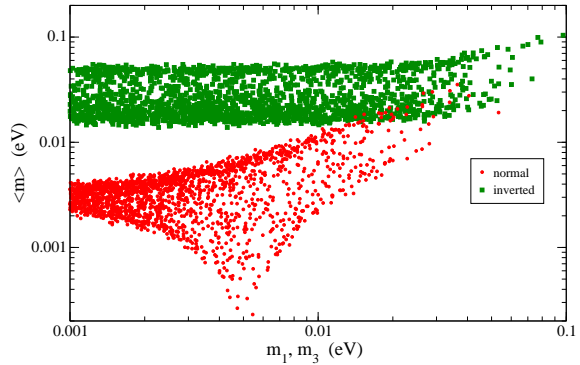}
\caption{Effective mass plot for neutrino-less double beta decay in 
  the case of perturbed tri-bimaximal mixing.}  
\label{fig:meff_brokenTBMa}
\end{figure}

It is clear that very accurate determinations of the neutrino mass 
hierarchy, the three mixing angles and eventually the three 
CP-violating phases will be required to pin down the most viable models.

\begin{theacknowledgments}

The author thanks Mu-Chun Chen and Werner Rodejohann for their 
contributions to parts of this survey, the organizers of the 
workshop for the opportunity to speak in the Neutrino Oscillation 
session, and the Fermilab Theoretical Department for its kind 
hospitality.  Fermilab is operated by the Fermi Research Alliance 
under Contract No. DE-AC02-07CH11359 with the U.S. Department of 
Energy.  
%
%

\end{theacknowledgments}

\end{document}